\documentclass[floatfix,aps,prl,twocolumn,groupedaddress,superscriptaddress,reprint]{revtex4-1}

\usepackage{graphicx}
\usepackage{dcolumn}
\usepackage{bm}
\usepackage{amsmath}
\usepackage{amssymb}

\RequirePackage[T1]{fontenc}
\RequirePackage[utf8]{inputenc}
\RequirePackage{lmodern}

\usepackage{siunitx}
\usepackage{braket}

\usepackage{isotope}

\newcommand{\UD}[0]{\ensuremath{\ket{\uparrow \downarrow}}}
\newcommand{\DU}[0]{\ensuremath{\ket{\downarrow \uparrow}}}
\newcommand{\singlet}[0]{\ensuremath{\ket{S}}}

\newcommand{\tzero}[0]{\ensuremath{\ket{T_0}}}
\newcommand{\bperp}[0]{\ensuremath{B^\perp_\mathrm{nuc}}}

\newcommand{\mr}[1]{\ensuremath{\mathrm{#1}}}

\newcommand{\ie}[0]{i.e.\ }

\begin{document}


\title{Measurement of back-action from electron spins in a gate defined GaAs double quantum dot coupled to a mesoscopic nuclear spin bath}


\author{P. Bethke}
\affiliation{JARA-FIT Institute for Quantum Information, Forschungszentrum J\"ulich GmbH and RWTH Aachen University, 52074 Aachen, Germany}
\author{R.P.G. McNeil}
\altaffiliation{present address: Center for Quantum Devices and Microsoft Quantum Lab Copenhagen, Niels Bohr Institute, University of Copenhagen, 2100 Copenhagen, Denmark}
\affiliation{JARA-FIT Institute for Quantum Information, Forschungszentrum J\"ulich GmbH and RWTH Aachen University, 52074 Aachen, Germany}
\author{J. Ritzmann}
\affiliation{Lehrstuhl f\"ur Angewandte Festk\"orperphysik, Ruhr-Universit\"at Bochum, D-44780 Bochum}
\author{T. Botzem}
\altaffiliation{present address: Centre for Quantum Computation and Communication Technology, School of Electrical Engineering and Telecommunications, UNSW Sydney, Sydney, New South Wales 2052, Australia}
\affiliation{JARA-FIT Institute for Quantum Information, Forschungszentrum J\"ulich GmbH and RWTH Aachen University, 52074 Aachen, Germany}
\author{A. Ludwig}
\affiliation{Lehrstuhl f\"ur Angewandte Festk\"orperphysik, Ruhr-Universit\"at Bochum, D-44780 Bochum}
\author{A.D. Wieck}
\affiliation{Lehrstuhl f\"ur Angewandte Festk\"orperphysik, Ruhr-Universit\"at Bochum, D-44780 Bochum}
\author{H. Bluhm}
\email[]{bluhm@physik.rwth-aachen.de}
\affiliation{JARA-FIT Institute for Quantum Information, Forschungszentrum J\"ulich GmbH and RWTH Aachen University, 52074 Aachen, Germany}
\email[]{bethke@physik.rwth-aachen.de}


\date{\today}

\begin{abstract}
Decoherence of a quantum system arising from its interaction with an environment is a key concept for understanding the transition between the quantum and classical world as well as performance limitations in quantum technology applications. The effects of large, weakly coupled environments are often described as a classical, fluctuating field whose dynamics is unaffected by the qubit, whereas a fully quantum description still implies some back-action from the qubit on the environment. Here we show direct experimental evidence for such a back-action for an electron-spin-qubit in a GaAs quantum dot coupled to a mesoscopic environment of order $10^6$ nuclear spins. By means of a correlation measurement technique, we detect the back-action of a single qubit-environment interaction whose duration is comparable to the qubit’s coherence time, even in such a large system. We repeatedly let the qubit interact with the spin bath and measure its state. Between such cycles, the qubit is reinitialized to different states. The correlations of the measurement outcomes are strongly affected by the intermediate qubit state, which reveals the action of a single electron spin on the nuclear spins.
\end{abstract}

\maketitle

The loss of quantum mechanical phase coherence of a system resulting from the interaction with some environment can in general be described by a full quantum mechanical model where system, environment and the coupling between them are represented by different terms of the complete Hamiltonian. In this setting, decoherence can be understood as averaging the system dynamics over fluctuations of the environment, or as entanglement between the two components.
This consideration implies that not only does the environment affect the system, but also that the system has a back-action on the environment. 
In the famous Born approximation holding if the system is coupled weakly to many degrees of freedoms of the environment, this back-action is neglected, thus assuming that the dynamics of the environment are independent of the state of the system and how it is manipulated. One may then model the effect of the environment as one (or several) classical noise fields, which greatly simplifies any computation. For example, assuming a stationary Gaussian process, all that needs to be known about the environment is its noise spectral density \cite{cywinski_how_2008}.
It is thus of great practical and fundamental interest to determine if the back-action of a controllable quantum system such as a qubit on its environment plays a role.

One suitable model for analyzing such scenarios is the central spin problem, where a central spin representing the quantum system under consideration is coupled to an environment of (many) other spins.
The central spin problem has been studied extensively in both theory \cite{de_sousa_theory_2003,barnes_nonperturbative_2012,witzel_quantum_2012,hanson_coherent_2008,stanek_dynamics_2013,cywinski_pure_2009,neder_semiclassical_2011,yao_theory_2006},  and
experiments in a wide range of physical systems \cite{witzel_electron_2010,pla_high-fidelity_2013,tyryshkin_electron_2012,bluhm_dephasing_2011,pla_single-atom_2012,malinowski_spectrum_2017,malinowski_notch_2017,laraoui_high-resolution_2013}.
For a small environment with strong coupling compared to other decoherence mechanisms, such as electron spins in nitrogen vacancy centers \cite{childress_coherent_2006,reinhard_tuning_2012} or donor sites in silicon \cite{madzik_controllable_2019} coupled to a few nuclear spins, it is possible to detect a back-action or even to transfer quantum states from the system to the environment and back \cite{taminiau_universal_2014}, which obviously requires a quantum description. 
Confined electron spins in quantum dots typically have a weaker coupling to a much larger nuclear spin environment of up to approximately $N = 10^6$ in gate-defined quantum dots in GaAs \cite{chekhovich_nuclear_2013}. 
A stronger back-action of the kind shown could be used to realize long-lived quantum memory in the nuclear spin bath \cite{taylor_long-lived_2003,gangloff_quantum_2019}.
Such devices thus constitute an interesting mesoscopic model system with a large but finite number of environmental spins.

A relatively simple example for a back-action that has been observed in this system is the polarization of nuclear spins from a repeated angular momentum transfer from the electron spin to the nuclei, referred to as dynamic nuclear polarization \cite{chekhovich_nuclear_2013,urbaszek_nuclear_2013}. Due to the (approximate) conservation of angular momentum, this polarization has a long lifetime and can therefore be built-up cumulatively over many cycles each comprising the initialization of the electron spin and subsequent interaction with the nuclear spins. Accordingly, it can be detected with many measurements taking in total much longer than any relevant dephasing time.
In contrast, back-action effects that occur on the time scale over which the electron spins lose phase coherence and potentially affect this dephasing process are much more subtle to detect.
In fact, all experiments on the dephasing and coherent dynamics of electron spins in gate-defined GaAs quantum dots to date in connection with nuclear spins can be described with semi-classical models in the above sense, i.e., by treating the nuclear spins as a classical variable not subject to any back-action \cite{bluhm_dephasing_2011, neder_semiclassical_2011, malinowski_notch_2017, pal_electron_2017}.

In this work we present direct experimental evidence for 
a  back-action from two coherently controlled electron spins in a gate-defined GaAs double quantum dot to their large nuclear spin environment that occurs during a single initialization-and-interaction cycle on the time scale of the electronic spin coherence time. It can thus be considered as a true quantum back-action, rather than an accumulative effect as encountered in dynamic nuclear polarization.
We conceptually follow the proposal of Fink and Bluhm \cite{fink_distinguishing_2014}, however leveraging $g$-factor anisotropy \cite{botzem_quadrupolar_2016} rather than quadratic coupling to transverse nuclear spins. The general idea is to use the correlations between the outcomes of two repetitions of a initialization-evolution-measurement (IEM) cycle that probes some bath property to reveal any interaction between qubit and bath \emph{in-between} the two cycles.

In a complete mathematical description, each IEM cycle would be represented by a positive operator valued measure (POVM) on the bath, causing an incomplete projection.
For a simpler intuitive picture, each IEM cycle can be thought of as a fully projective measurement. 
The first measurement would then project the bath into some eigenstate or eigenspace.
Neglecting internal bath dynamics for simplicity, the second measurement would measure the same corresponding eigenvalue. To test if the qubit influences the bath dynamics, the qubit is prepared in different states in-between the two measurements. Any back-action that changes the bath properties probed by the IEM cycles then affects the correlations between the measurement outcomes. Due to the correlation approach, the scheme can be applied with arbitrary initial states of the environment, including fully mixed states as encountered in the system studied here.

\begin{figure}[!t]
\centering
\includegraphics{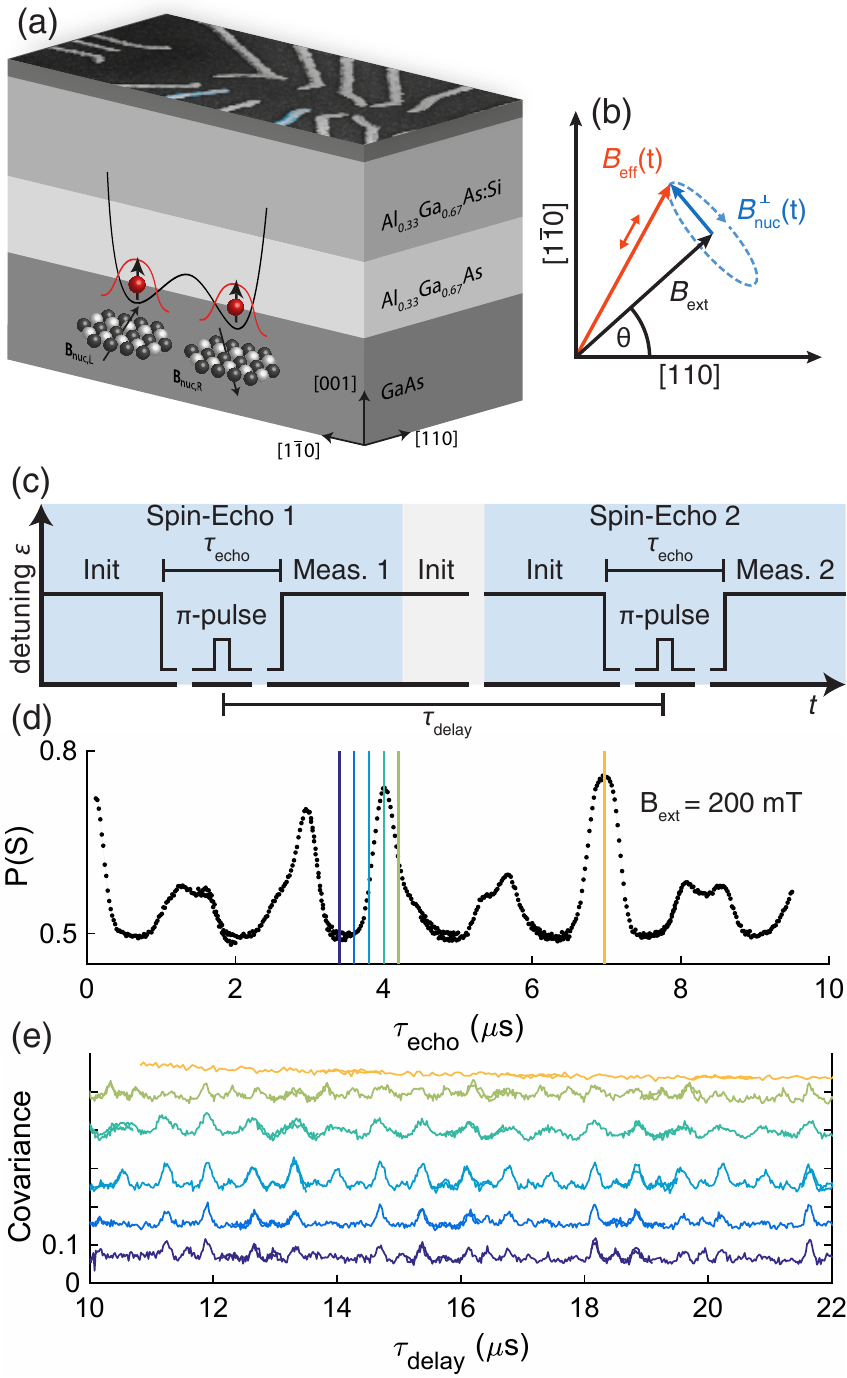}
\caption{\label{fig:fig1}
\textbf{(a)}
    The double quantum dot is formed in a 2DEG inside a GaAs/AlGaAs heterostructure by electrostatic gating from the top.
\textbf{(b)}
    If the external magnetic field $B_\mathrm{ext}$ is not aligned with the $[1 1 0]$ or $[1 \overline{1} 0]$ crystal axes, $g$-factor anisotropy leads to a misalignment of the electronic and nuclear quantization axes, which can be modeled as a linear coupling from \bperp{} to the magnetic field $B_\mathrm{eff}$ that the electrons see.
\textbf{(c)}
    Two individual spin-echo experiments are carried out one after the other and the correlation of their outcomes is calculated numerically from many repetitions. In-between, the qubit is reinitialized to the $S$ ground state.
\textbf{(d)}
    Singlet return probability of a spin-echo experiment (blue block in (c)). The revivals correspond to echo evolution times where $\tau_\mathrm{echo} / 2$ is a multiple of two or three nuclear Larmor periods. Valleys indicate sensitivity to the nuclear Larmor phase.
\textbf{(e)}
Covariance of two spin-echo measurements as a function of the time $\tau_\mathrm{delay}$ between them, using the same color code as the vertical lines in (d) to indicate the evolution time $\tau_\mathrm{echo}$. The traces are offset by \num{0.1} for clarity (tick marks). 
}
\end{figure}

The qubit studied here is a two-electron spin qubit. The electrons are confined in a double quantum dot formed inside a two-dimensional electron gas in a GaAs/AlGaAs heterostructure by electrostatic gating (see Fig.\,\ref{fig:fig1}(a)).
The qubit is encoded in the $m_z=0$ subspace spanned by the singlet $\singlet = (\UD{} - \DU{}) / \sqrt{2}$ and the triplet $\tzero = (\UD{} + \DU{}) / \sqrt{2}$ states, where the first(second) arrow indicates the electron spin state in the left(right) dot.
The remaining two states, $\ket{T_+} = \ket{\uparrow \uparrow}$ and $\ket{T_-} = \ket{\downarrow \downarrow}$, are split off by the Zeeman energy $ E_z = g^* \mu_B B_{\mathrm{ext}}$ induced by an external magnetic field. The use of two spins facilitates manipulation of the qubit via the exchange interaction, but the observed effects apply equally to single electron spins.
The exchange coupling is varied by rapidly pulsing the detuning $\varepsilon$, which is the energy difference between the left and right quantum dot. 
The $\UD$ and $\DU$ states with one electron in each dot can be prepared by adiabatically sweeping the detuning between the two dots, thereby separating the electrons.
All experiments were carried out in an external magnetic field of \SI{200}{\milli \tesla}, which achieves a rough balance between long coherence time and sensitivity to transverse nuclear fields of the Hahn-echo experiments \cite{botzem_quadrupolar_2016} and is likely not very specific.

In the well-separated charge state with one electron in each dot and for large external magnetic field $B_\mr{ext}$, the dynamics of the $\singlet$-$\tzero$ qubit can be modelled using the hyperfine Hamiltonian (supplement of Ref. \cite{botzem_quadrupolar_2016},\footnote{Note that Ref. \cite{botzem_quadrupolar_2016} uses two inconsistent sign conventions, which are of no physical consequence.})
\begin{align}
    \label{eq:hamiltonian}
     \hat{H} = \mu_\mathrm{B} g_\parallel \left( B_\mr{ext} + B^\parallel_\mr{nuc} - \frac{g_\perp}{g_\parallel} B^{\perp}_{\mr{nuc}}(t) + \frac{B^\perp_{\mr{nuc}}(t)^2} {2B_\mr{ext}} \right) \hat{S}_z
\end{align}
for each dot, where $\mu_\mr{B}$ is the Bohr magneton, $g_{\parallel,\perp}$ are entries of the anisotropic electronic $g$-tensor, $\hat{S}_z$ is the electron spin operator, and $B_\mr{nuc}^{\parallel,\perp}$ are the parallel and perpendicular components of the Overhauser field. The linear coupling to $B_\perp$ exploited here is caused by a $g$-factor anisotropy of about \SI{5}{\percent} causing different quantization axes for electron and nuclear spins \cite{botzem_quadrupolar_2016} if the external in-plane magnetic field is applied along the $[1 0 0]$ crystal axis.
$B^{\perp}_\mr{nuc}$ is the sum of three components each of which oscillates at the nuclear Larmor frequency of the corresponding nuclear isotope in GaAs. The other terms are quasi-static on a qubit timescale. The last term quadratic in $\bperp$ is not relevant for our purpose.

For each of the IEM cycles, we chose a Hahn-echo dynamical decoupling sequence as shown in the blue blocks in Fig.\,\ref{fig:fig1}(c). The qubit is initialized in a $\singlet$ state, evolves under the influence of the nuclear spin bath with exchange coupling off except for the $\pi$-pulse, and is finally measured in the singlet-triplet basis. The echo-pulse cancels out the influence of all static terms on the electron spin \cite{bluhm_dephasing_2011}.
The phase picked up by the qubit during the free evolution in the spin-echo sequence due to the time-dependent $B^{\perp}_\mr{nuc}(t)$ is generally not eliminated. Fig\,\ref{fig:fig1}(d) shows the singlet return probability of a single cycle as a function of the echo time. By repeating the experiment many times with a random initial nuclear spin configuration each time, the results are averaged over the initial phase of the nuclear Larmor precession, i.e., the direction of the total transverse Overhauser field. On the peaks, the evolution time $\tau_\mr{echo} / 2$ is a multiple of the nuclear Larmor period, so the net qubit phase is zero regardless of the initial nuclear Larmor precession phase. 
The valleys occur at times $\tau_\mr{echo}$ where the singlet return probability is most sensitive to the initial phase, so that averaging over it leads to a complete suppression of the echo.
Each echo measurement therefore constitutes a measurement of the nuclear Larmor phase of $B^{\perp}_\mr{nuc}(t)$.
The complicated peak structure is a result of the presence of three nuclear spin species with nearly commensurable Larmor frequencies \cite{botzem_quadrupolar_2016}, and can be understood by applying the argument to each species separately.

To reveal the internal spin bath dynamics we investigate the correlations using the covariance
\begin{align}
C = \left\langle x_1 x_2 \right\rangle - \left\langle x_1 \right\rangle \left\langle x_2 \right\rangle
\end{align}
of the outcomes $x_i$ of two subsequent Hahn-echo measurements (labeled Meas. 1 and Meas. 2) by averaging over the single-shot results of many repetitions of the sequence in Fig.\,\ref{fig:fig1}(c).
$C$ depends on the echo time $\tau_\mathrm{echo}$ and the time between echo $\pi$-pulses $\tau_\mathrm{delay}$. To probe this dependence, the qubit is reinitialized and locked to the $\singlet$ ground state with both electrons in the same dot in-between the two IEM cycles to minimize the hyperfine interaction between the electron and nuclear spins.

If the free evolution time $\tau_\mathrm{echo} / 2$ of both echo measurements is a multiple of the nuclear Larmor periods of all three spin species in the host crystal, the echoes are insensitive to the phase of the nuclear precession, leading to the revival of the echo response at  $\tau_\mr{echo} = \SI{7}{\micro\second}$ in Fig.\ref{fig:fig1}(d). As a result, the covariance between both echo measurements becomes independent of the delay between them, as can be seen from the corresponding curve in Fig.\ref{fig:fig1}(e).
For other $\tau_\mathrm{echo}$, the covariance reveals the nuclear Larmor
precession dynamics of the nuclear spin bath. It peaks if the time between echoes $\tau_\mathrm{delay}$ matches a multiple of the Larmor periods. The visibility of the three nuclear spin species strongly depends on $\tau_\mathrm{echo}$.
This correlation measurement scheme reveals the nuclear Larmor precession in a similar way to previous work which investigated nuclear spin dynamics using correlations between Landau-Zener sweeps across the $S$-$T_+$-transition \cite{nichol_quenching_2015,dickel_characterization_2015, pal_electron_2017}.

\begin{figure}[!t]
\includegraphics{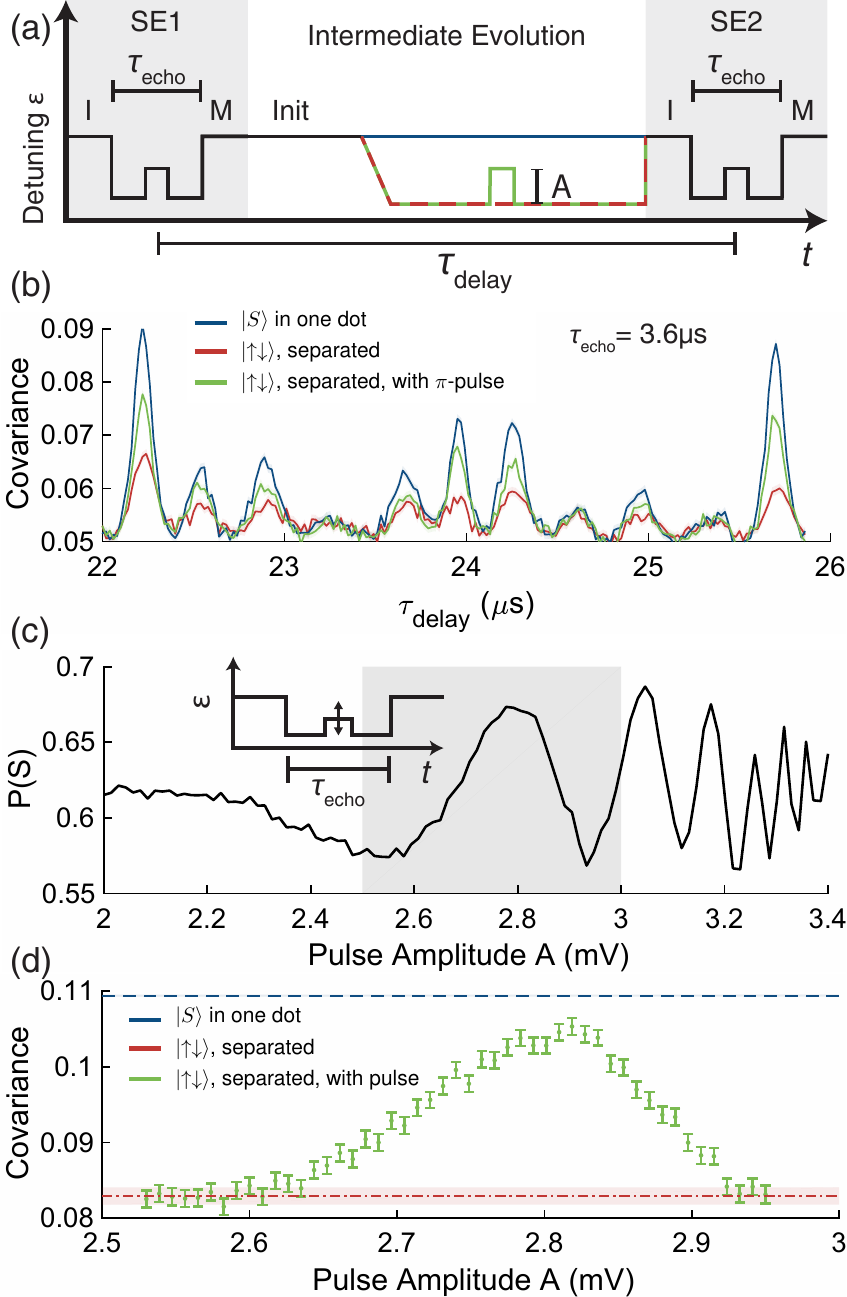}
\caption{\label{fig:fig2}
Measurements revealing back-action
\textbf{(a)}
Schematic of the pulse sequences. In-between two spin-echo sequences and measurements SE1,2, we prepare the qubit electrons in different states: singlet at a bias point with large exchange splitting (blue line), adiabatically prepared $\UD{}$ with separated electrons (red dashed line) and separated \UD{} with an exchange $\pi$-pulse after half the evolution time (green line).
\textbf{(b)}
Covariance obtained for these intermediate sequences. We observe the maximum amplitude for the $\singlet$ state which induces no Knight shift. In the \UD{} case, the amplitude is reduced because the two nuclear spin baths experience a Knight shift of opposite sign. Some of the amplitude can be recovered if the electrons are exchanged $\UD{}$ to $\DU{}$ after half the evolution time by a $\pi$-pulse, canceling out the Knight shift.
\textbf{(c)}
Dependence of the singlet return probability of a single Hahn-echo sequence on the amplitude of an exchange pulse.
The first maximum corresponds to a $\pi$-rotation, the inset shows the corresponding pulse sequence.
\textbf{(d)}
Covariance as a function of the exchange pulse amplitude of the intermediate pulse for a fixed time $\tau_\mathrm{delay} = \SI{22.3}{\micro \second}$.  The agreement with panel (c)(shaded region) shows that the correlations indeed depend on the electron state during the intermediate evolution, thus confirming a back-action of the electron spin state.
}
\end{figure}

To examine the effect of different intermediate qubit states, we prepare the qubit in different states with separated electrons in-between the IEM cycles, instead of keeping both electrons in the same dot in the $m_z=0$ ground state.
If the electrons are prepared in the long-lived $\UD$ state with separated electrons, the covariance is greatly diminished.
In this case, the nuclear spins coupled to each dot experience a Knight shift of opposite sign due to the
electron spin, \ie{}, a different magnetic field. Therefore the nuclear spins in both baths precess at different speeds and pick up a different Larmor phase, leading to a reduction in peak height in the covariance measurement (see Fig.\,\ref{fig:fig2}(b)) compared to the case where the electrons are prepared in the same dot.
To verify that this loss in covariance is indeed due to the back-action of the electron spin, we
perform an exchange $\pi$-pulse after half the intermediate evolution time, changing
the $\UD{}$ electron state into the \DU{} state. Fig.\,\ref{fig:fig2}(b) shows that the peak height is partially restored, because the Knight shift in the first
and second half of the intermediate evolution cancel each other out to some degree. The remaining loss is likely a result of imperfections in the exchange
pulse, for example due to charge noise, calibration errors, or non-zero $\Delta B_\mr{nuc}^\parallel$ \cite{botzem_quadrupolar_2016}.

To further substantiate that the observed covariance restoration is really due to the electron state and not some direct effect of the driving pulse, we fix the echo time $\tau_\mathrm{echo}$, the time between echoes $\tau_\mathrm{delay}$, and the exchange pulse duration during the intermediate evolution and vary the exchange pulse amplitude. The amplitude of the $\pi$-pulse in the outer echoes remain unchanged.
The result is shown in Fig.\,\ref{fig:fig2}(d) and compared to the two other intermediate sequences (dashed lines): Singlet preparation ($m_z = 0$) as well as  adiabatic preparation of the \UD{} electron state as discussed above. 
The resulting variation of the correlation strength closely resembles the singlet return probability seen in single echo experiments with varied pulse amplitude. The oscillatory behavior reflects  the induced rotation on the Bloch sphere (cf.\,Fig.\,\ref{fig:fig2}(c)).
If the pulse is a perfect $\pi$-pulse, the qubit is in the \DU{} state in the second half, otherwise in some superposition given by the rotation angle. We observe that the covariance closely follows the echo singlet return probability in Fig.\,\ref{fig:fig2}(d), continuously connecting the maximum expected value set by an intermediate singlet state with both electrons in one dot and the minimum set by the $\UD$ intermediate state with separated electrons. This observation demonstrates that the reduction of correlation is indeed dependent on the electron state, thus confirming the presence of a back-action.
The absolute covariance values differ slightly from the values in Fig.\,\ref{fig:fig2}(b) because the data was taken in separate experiments.

In summary, we implemented a general scheme proposed by Fink and Bluhm \cite{fink_distinguishing_2014} for the detection of quantum-mechanical back-action in a qubit-bath system by observing how correlations between two consecutive initialize-measure-evolve (IEM) cycles  vary depending on how the qubit is manipulated in-between them. We unambiguously detect a back-action from single electron spins to a mesoscopic environments of about $10^6$ nuclear spins on a timescale that is comparable to the coherence time of both the qubit and nuclear spins. 

Note that the environment can be considered ergodic because the experiment wall time is much longer than the autocorrelation time of the slowest degree of freedom, namely the longitudinal Overhauser field with an autocorrelation time on the order of seconds to minutes \cite{tenberg_narrowing_2015, reilly_measurement_2008}. While a single repetition of the experiments in Fig.\,\ref{fig:fig2}(d), for example, takes \SIrange{50}{100}{\micro \second} each point is an average over a total of around \SI{250}{\second} measurement time spread out in an outer loop across \SI{15}{\hour} of wall time. Thus, the nuclear spin environment can be well described by a fully mixed state.

While for a pure bath state, decoherence is always associated with entanglement, the situation is more subtle for such a mixed state.
It was shown in Ref.\,\cite{roszak_characterization_2015} that for a pure dephasing Hamiltonian as considered here, a dependence of the bath evolution on the qubit state, as observed in our experiments, implies that the qubit and the bath become entangled. 
 This conclusion is particularly interesting in the light that the bath starts out in a maximally mixed state, which implies that no entanglement can be generated by a single interaction alone \cite{sakuldee_characterization_2019}.
Here, the first interaction and measurement produces a nontrivial bath state, so that the second coupled evolution can indeed generate entanglement.

While the separation time in our experiments is significantly longer than the duration of individual qubit gates, \SI{20}{\micro\second} compared to \SI{20}{\nano\second}, it is on the same order as the coherence time of the qubit under dynamical decoupling \cite{bluhm_dephasing_2011, malinowski_notch_2017} of around \SI{100}{\micro\second}. Therefore the effect is likely not relevant for the fidelity of individual gates, but may affect algorithms or quantum error correction schemes composed of many gates.
Our scheme can also be employed to probe back-action effects in other types of qubits with varying coupling and number of degrees of freedom in the environment to examine the quantum-classical transition.

The observed back-action is closely related to the uncertainty relation so that the experiment represents a test of one of the fundamental principle of quantum mechanics. The three evolution cycles can be thought of as measurements of the environment. If the operators corresponding to the outer and intermediate measurement do not commute, an uncertainty relation between them forbids a sharp joint measurement. This effect arises from the back-action of the intermediate measurement bringing the environment out of an eigenstate of the outer measurement even if it was prepared in such a state by the first one. A failure to detect a back-action that should be visible according to a well-calibrated quantum-model could thus point to a violation of
a central prediction of quantum mechanics. In the present case, such a qualitative discrepancy is not observed. Comparison with a sufficiently accurate model of the experiment could lead to a quantitative test.


\begin{acknowledgments}
This project received funding from Deutsche Forschungsgemeinschaft under Grant No. BL 1197/4-1, the Alfried Krupp von Bohlen and Halbach Foundation, and the Excellence Initiative of the German federal and state governments.
J.R., A.L., and A.D.W. acknowledge gratefully support of DFG-TRR160,  BMBF--Q.Link.X  16KIS0867, and the DFH/UFA  CDFA-05-06.
We acknowledge support by the Helmholtz Nano Facility (HNF) at the Forschungszentrum J\"ulich \cite{albrecht_hnf_2017}.
We thank Federica Haupt for support in compiling the paper.

\section{Author contributions}
Molecular-beam-epitaxy growth of the sample was carried out by J.R., A.L. and A.D.W.. T.B., P.B., and R.P.G.M. set up the experiment. R.P.G.M. fabricated the sample. P.B. conducted the experiment. P.B. and H.B. analyzed the data and prepared the manuscript.
\end{acknowledgments}


%

\end{document}